\documentclass{article}  
\usepackage{lajolla2006}
\usepackage{graphicx}
\frompage{000} \topage{000}                                              
\def\rightmark{sQGP formation time}
\def\leftmark{V.S Pantuev}
 
\title{Evidence of finite sQGP formation time at RHIC } 
\authors{
{V.S Pantuev %
}\\[2.812mm]
{\normalsize
\hspace*{-8pt}University of Stony Brook,  
Stony Brook, NY 11720, USA\\[0.2ex] 
}}
 
\abstract{
We demonstrate that the existence of a finite formation time of strongly interacting plasma in nuclear collisions 
at RHIC is an inavitable conclusion from  recent experimental data. 
The most striking feature of the experimental data - an absense of absorption of high transverse momentum pions 
in the rection plane direction 
for mid-peripheral collisions - points to the presence of a surface zone with no absorption 
and strong suppression in the inner core. A natural interpretation 
of such a zone should be the plasma formation time $T$$\simeq$ 2-3 fm/c. 
We discuss constraints induced by finite formation time on some physical observables at RHIC. 
Nuclear modification factor, azimuthal assymetry, di-jet correlations can quantitatively be described 
by particle production in the early stage of the nuclear collision. A possible impact on azimuthal anisotropy 
at lower hardon momenta, on interpretation 
of the non-photonic electrons and J/$\psi$ data is also considered.
}

 
\begin{document}
 
\maketitle
\setcounter{page}{1}

\section{Introduction}\label{intro}
Nowadays, the physics of relativistic nuclear collisions at RHIC goes from the phase of discoveries~\cite{whitepaper} to
the stage of understanding properties of a new state of nuclear matter: the quark-gluon plasma, QGP. 
Naive expectations that QGP to be a gas of weakly interacting quark and gluons was washed out by very first experimantal 
data. Collective phenomena, such as radial and elliptic flow, reveal ``hydro'' properties of the QGP, which behave 
``like a good liquid rather than a dilute gas of quasiparticles''~\cite{shuryakzahed}. The idea was introduced, that 
plasma at RHIC should be in strongly coupled regime, sQGP. Another phenomenon seen at RHIC is {\it jet quenching}. Even 
very hard jets are expected to lose some energy in the formed medium. The quenching is decribed by a nuclear 
modification factor, $R_{AA}$, which is defind as the observed number of jets normalized to the expected number of jets 
from superposition of individual nucleon-nucleon collisions. Very first RHIC data have indeed shown significant particle 
suppression at high $p_T$~\cite{whitepaper}. 
Recently, the PHENIX collaboration 
showed a preliminary result for the dependence of $R_{AA}$ on the azimuthal angle $\phi$ relative to the reaction plane in Au+Au 
collisions at $\sqrt{s_{NN}}$=200 GeV at RHIC~\cite{cole,david,winter}, Fig.~\ref{fig:Raa_vs_phi}.
The most interesting feature of these data is that, at event centrality class 50-60$\%$, 
for transverse momenta above 4 GeV/c, 
in-plane $R_{AA}$ equals to one within the errors. 
This implies no absorption at all for high $p_T$ pions. For such high momenta, Cronin effect~\cite{cronin} is 
negligible and can not bring $R_{AA}$ close to one. 
At the same time, a significant particle 
absorption is seen in out-of-plane. 
Apparently, parton energy loss calculations can't describe this ``punch through'' feature of in-plane $R_{AA}$=1, e.g. see  
Fig.5 in Ref.~\cite{david}.
In our resent papers~\cite{pantuev1,pantuev2} we offer an alternative explanation of this and certain other features of the data. 
We introduce a new term {\it plasma formation time}, which could be considered as a transision time for produced matter 
to reach liquid or strongly interacting phase. During this time, soft and hard jets can escape from interaction zone 
with very little or no energy loss. Only jets originating on the surface can reach the detector. It forms kind of corona around nuclear 
overlap region with thickness significantly larger than nuclear diffuseness zone.  
In this paper we describe consequences of sQGP formation time on some physical observables.  
 
\begin{figure*}[hbt]
\begin{minipage}[t]{0.8\linewidth}
\includegraphics[width=1.0\linewidth]{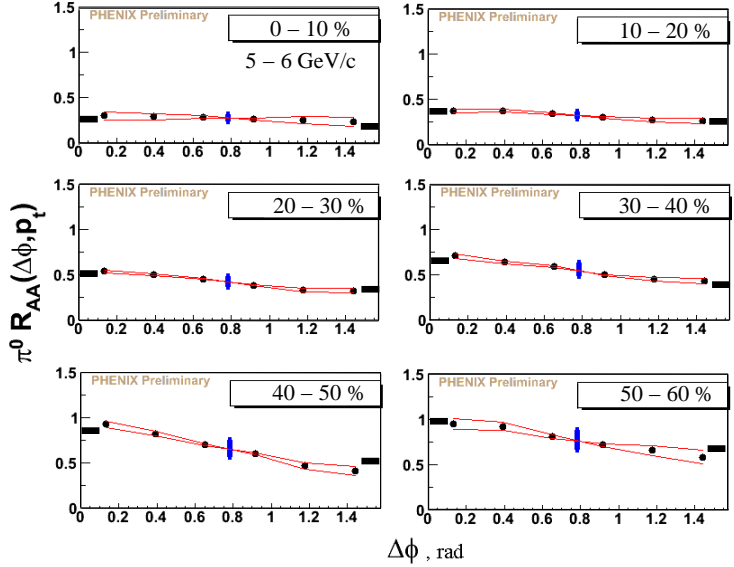}
\end{minipage}
\label{fig:Raa_vs_phi}
\caption{\label{fig:Raa_vs_phi} PHENIX preliminary  $R_{AA}$ results for $\pi^0$ at momenta 5-6 GeV/c 
versus angle $\phi$ relative 
to the reaction plane for different centralities in Au+Au collisions at RHIC~\cite{winter}. Points are experimental data, 
thin lines show systematic errors from the reaction plain resolution, vertical bars in the middle show avaraged 
over the reaction plane $R_{AA}$ value and it's error.
Black horizontal bars are our predictions.}
\end{figure*}

\section{Where formation time comes from?}\label{model}  

\begin{figure}[thb]
\includegraphics[width=0.6\linewidth]{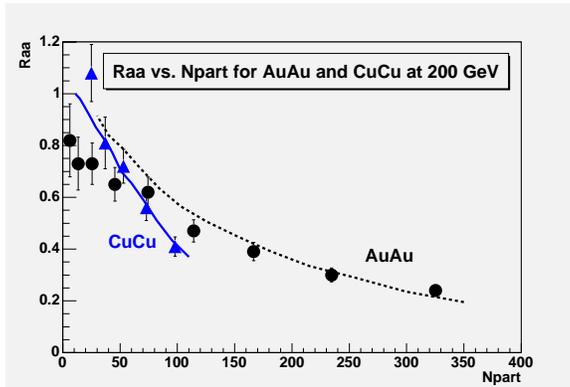}
\caption{\label{fig:Cu}  $R_{AA}$ for Au+Au (dashed curve) and Cu+Cu (solid curve) collisons versus the number of
participant nucleons, $N_{part}$. The cicles are experimental $\pi^0$ data for Au+Au collisions integrated for 
$p_T$$\ge$4 GeV/c~\cite{pi0}. The triangles are data for Cu+Cu collisions of  
$p_T$$\ge$7 GeV/c~\cite{maya}. Only statistical errors are shown.}
\end{figure}

Let me briefly describe the arguments were used in Ref.~\cite{pantuev1}. 
We construct a simple model using the Monte Carlo simulation of nucleus-nucleus collisions based on 
the Glauber approach and discussed in detail in~\cite{jjia}. A Woods-Saxon density distribution was used in 
all cases.  
We restrict ourselves to describe the data of high $p_T$ pions with transverse momentum above 4 GeV/c, where $R_{AA}$ 
does not depend on  $p_T$. 
We assume that all high $p_T$ pions are produced by parton fragmentation and that the  number of partons is proportional 
to $N_{binary}$.
If there is no absorption,  
$R_{AA}$=1 in all directions and the shape of the event is isotropic. 
To explain the experimentally observed feature of in-plane $R_{AA}$$\simeq$1, we investigate the role of
purely geometrical factors. 
In our model, jets which have to travel through the medium at $some$ $direction$ from thier production point to 
the surface less than a distance $L$ 
will leave the interaction 
zone unmodified. Jets originating in the core region deeper than $L$ suffer significant 
energy loss and are completely absorbed. 
We can expalin the picture in other words: let for each jet to be  free streaming at $any$ direction with speed of 
light, $c$, for some time $T$. After that time we stop all jets which still left within the envelop of Woods-Saxon 
radia of overlapping nuclei.     
The whole picture looks like a pure $corona$ jet production, 
but we allow this corona region to be larger than a  Woods-Saxon type skin. 
 We can calculate $R_{AA}$ as
a ratio of the $seen$ collisions, $N_{coll}$, to the total $N_{binary}$ at particular centrality class.
The cut-off 
parameter $L=T*c$ should 
be on the  order of the size of the in-plane interaction zone at 50-60$\%$ centality. 
Parameter $L$ was found to be $L$=2.3$\pm$0.6 fm to get 
$R_{AA}^{in}$=0.9$\pm$0.1 for 50-55$\%$ centrality. This is close 
to what is seen experimentally and  
leaves some room for Cronin enhancement~\cite{cronin}, if any.
The results of our calculations together with experimental data for Au+Au collisions at 200 GeV 
are shown in Fig.~\ref{fig:Raa_vs_phi} and Fig.~\ref{fig:Cu}. In Fig.~\ref{fig:Cu}, we also plot our 
results for $R_{AA}$ at high $p_T$ in Cu+Cu collisions at 200 GeV using the same $L$=2.3 fm. 
There is an axcellent agreement with all experimental data.

We investigated the sensitivity of our result to various assumptions, like:  
if the thickness of the material integrated over the path length is a critical 
cut-off parameter, the centrality dependence becomes very strong in this case and 
can not describe the data. 
Another premise tested uses the number of participant nucleons,  $N_{part}$, instead 
of $N_{coll}$. In this case the centrality dependence is weaker than in the experimental data. 
To take into account obvious fluctuations in the jet absorption we smooth the cutoff edge $L$ by introdusing 
its diffuseness $D$ of 0.1-2 fm. The results do not change significantly 
in this parameter range. A maximum deviation of 5$\%$  was obtained for $D$=2 fm,  
only in very peripheral collisions. The default value was choosen to be $D$=1 fm. \\ 
  Within our model we 
can also describe $R_{AA}$=0.36 for neutral high $p_T$ pions in Au+Au collisions at $\sqrt{s_{NN}}$=62.4 GeV~\cite{david} 
in most central events as measured by PHENIX, but with even larger value $T$=3.5 fm/c. 

\section{Discussion and Conclusions}\label{discussion} 
What could be the physical interpretation of the cutoff  $T=L/c$? Our guess is that 
it is natural to assign this parameter $T$ to a 
{\it plasma formation time}, or, at least,  the time when parton energy loss actually starts. 
This gives a simple and elegant interpretation of the effect: 
particles produced close to the surface of the collision zone have time, about 2-3 fm/c, to 
escape. After that time a dense and strongly interacting matter, sQGP, is  
formed and  high  $p_T$ partons suffer from a significant energy loss. This formation time puts 
significant constraints on many ``explained'' and not explained yet observables at RHIC. 
Let us go through the list of such observables and some physical consequences.
\begin{figure}[thb]
\includegraphics[width=0.6\linewidth]{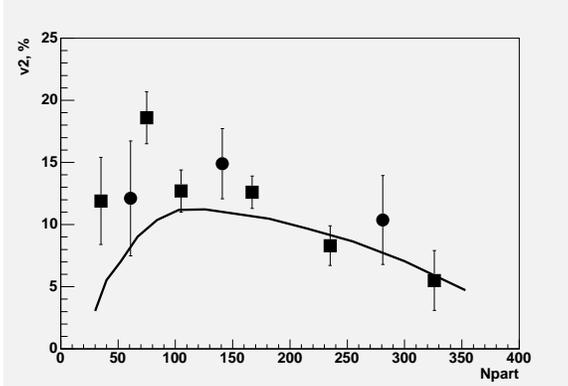}
\caption{\label{fig:v2} Calculated ellipticity parameter $v_2$ for Au+Au collisons, solid line, versus the number of
participant nucleons, $N_{part}$. Data for $\pi^0$ with error bars are: cicles for 4.59 GeV/c, 
squares for 5-7 GeV/c. 
PHENIX preliminary data~\cite{kaneta,winter}.}
\end{figure}

\begin{figure}[thb]
\includegraphics[width=0.6\linewidth]{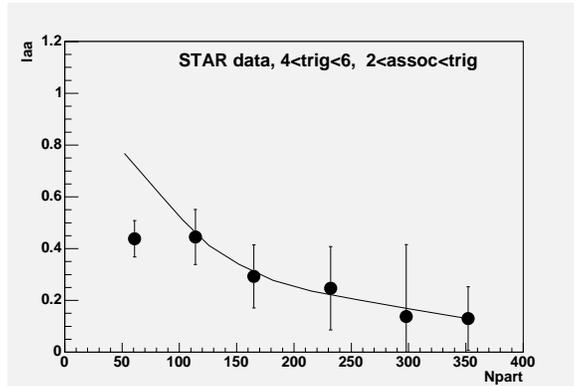}
\caption{\label{fig:Iaa}  Calculated $I_{AA}$ for Au+Au, the curve, versus the number of
participant nucleons, $N_{part}$. The width (sigma) of away-side jet is 0.5 radians.
The cicles are experimental data from~\cite{dijets}. }
\end{figure}

1) Formation time serves as a {\it corona} for particle production at the early stage of the collision. 
So, we have to split particle production into two sources, from the core and from the corona. Most of soft 
particle are coming from core, all hard particles are produced in corona. There should be the later 
stage hadron gas corona too, but it might have very little influence on high $p_T$ particles.\\
2) The core region is very opaque, {\it all jets} are stopped in the core. Indeed, there are some theoretical 
estimations that jets would not make it beyond 1/3 or 1/2 fm~\cite{shuryakzahed2,zahed}.\\
3) All pions with momentum above 3-4 GeV/c, all baryons - above 5-6 GeV/c are coming purely from corona,
this is why $R_{AA}$ is constant at high $p_T$~\cite{pi0}. It also means that the information about 
sQGP properties lays solely on low momenta particles.\\
4) The experimental variable, $v_2$, describes the elliptic shape of the event in the nucleus-nucleus
collision in azimuthal angle. We estimate $v_2$ for high $p_T$ particles by the jet survived 
probability in and out of plane, Fig.~\ref{fig:v2}. The parameter $v_2$ reaches 11-12$\%$ in mid-central events (centrality 30-35$\%$)  
and nicely follows the trends observed in the experiments~\cite{kaneta,winter}. Our result 
disproves the assertion 
 that jet quenching models can not explain the measured $v_2$ at high $p_T$~\cite{shuryak}. 
It also means that measured $v_2$ at high $p_T$ is not sensitive to hydro/collective flow.\\
5) Combining statements 1 and 4 we come to the conclusion that at intermediate momentum range, 2-4 GeV/c, 
particles from early corona with the power law spectrum shape and with the smaller value of $v_2$ dilute  
the exponentially falling spectrum from the core with hydro-like $v_2$ value. This is why $v_2$ at momenta 
below 2 GeV/c rises with close to hydro calculation values, saturates at some value, and then starts to fall down~\cite{flow} 
to the values we present in Fig.~\ref{fig:v2}. Would we develope the method to subtract corona contribution 
from the observed $v_2$ value, the agreement with hydro calculation will be even better. At intermediate 
$p_T$ region mesons are much more suppressed than baryons, this is why mesons azimuthal anisotropy goes down 
earlier than for baryons and, so called, constituent quarks $v_2$ scaling~\cite{flow} could be accidental.\\
6) Undistorted production of high $p_T$ particles from the corona region explains 
the lack of change of  
hadron distributions in forward jets with  centrality and orientation relative to the reaction plane~\cite{dijets}.\\
7) The same is true for di-jets. The yield of high $p_T$ di-jets will be suppresed, leading dominantly to the 
tangential  di-jet prodiction with no change of jet properties. 
STAR collaboration published the results on $I_{AA}$ in Au+Au collisions - the ratio of away-side yield per 
trigger particle to the similar value from p+p collisions~\cite{dijets}. We calculate $I_{AA}$ within our model 
with $T=2.3 fm/c$ for different centralities. The increase of corona thickness by 2.3 fm keeps the value of 
$I_{AA}$ at the level close to the experimental data, Fig.~\ref{fig:Iaa}.\\

\begin{figure}[thb]
\includegraphics[width=0.6\linewidth]{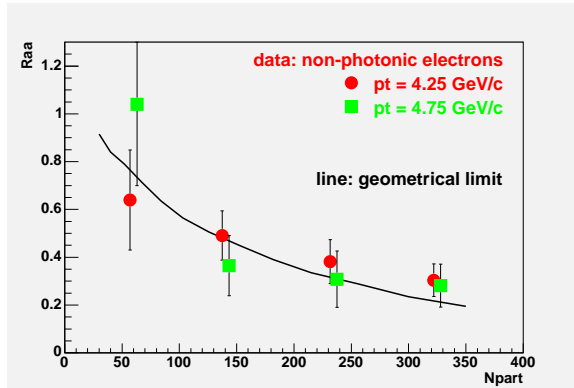}
\caption{\label{fig:charm} $R_{AA}$ for  non-photonic electrons in Au+Au versus the number of
participant nucleons, $N_{part}$. The PHENIX preliminary experimental data are from~\cite{charm}: circles and squares are for 
$p_t$=4.25 GeV/c and 4.75 GeV/c, respectively. Shown are statistical errors only. 
Solid line - the result of our calculation which puts the geometrical limit in case of strong charm quark absorption.}
\end{figure}

8) In Fig.~\ref{fig:charm} we present PHENIX preliminary results on non-photonic electrons at high momenta 
shown at Quark Matter 2005~\cite{charm}. We also plot the lower limit of $R_{AA}$ determined by 
particle production from the  corona region during the formation time, solid line. Experimental points sit exactly on 
the line. These electrons originated from D-mesons, charmed baryons and, probably, from 
bottom quark decays. The conclusion of Fig.~\ref{fig:charm} would be that at high $p_t$ the charm quarks are completely 
absorbed by
dense medium, similar to the light quarks.\\
9) In the case of a strong charm quark absorption, following the argument in our previous part 4, one can 
expect a significant $v_2$ for non-photonic electrons. Experiment indeed found non-zero $v_2$~\cite{charm}.\\
10)  Many interesting features may come from J/$\psi$ production in nuceus-nucleus collisions. 
 In our consideration we devide nucleus-nucleus collision in to two stages: before plasma formation and 
after. So, it should be different  J/$\psi$s absorbtion processes at each stage of the collision. If in some 
momentum range J/$\psi$s are absorbed by the core, there should be non-zero $v_2$ too.\\
11) Few speculative conclusions could be made from the beam energy dependence of $T$. At smaller beam 
energy the measured formation time rises, at 200 GeV $T$=2.3 fm/c, at 62 GeV $T$=3.5 fm/c. At lower energy $T$ should be 
 even larger, such that sQGP can't be formed at all because of fast longitudinal expansion. Indeed, 
at 20 GeV, see talk by C.~Klein-Bosing at this workshop,  $R_{AA}$=1. Such behaviour goes in favor of 
the existence of strongly coupled  3- and 4-body bound states in sQGP phase, as it was proposed by Edward Shuryak and 
Ismail Zahed~\cite{shuryakzahed}. Time, necessary to form such states, should be proportional to 
the mean distance between interaction points with color exchange. This distance, itself, should be inversally proportional to 
the square root of density of such interactions. {\it If relativistic rise} of the total nucleon-nucleon cross section 
comes purely from the parton hard scatterings, we can estimate the relative contribution of hard scatterings in 
the total number of nucleon-nucleon collisions. At 20 GeV the NN total cross section is at its minimum  - 
there is no hard scattering. At 62 GeV cross section rises by 5 millibarns, at 200 GeV - by 13 mb, at 5500 GeV - 
by 49 mb. Formation time of bound states should be {\it proportional} to one over the square root of these numbers. 
If we measure $T$=2.3 fm/c at 200 GeV, from rise of the NN total cross section, we estimate $T$=3.6 fm/c at 62 GeV, 
$T$=1.2 fm/c at 5500 GeV, LHC energy.\\

12) {\it  The existence of formation is a direct sign that sQGP is actually formed at RHIC.}\\

We present an $orthogonal$ point of view on some experimental 
data. This view could be useful for better understanding of properties of a new matter produced 
at RHIC. Perhapes, it will bring clear vision or some new ideas on formation process and properties 
of a quark-gluon phase. 

\section*{Acknowledgments}
This work was partially supported by the US-DOE grant DE-FG02-96ER40988.

\vfill\eject
\end{document}